\documentstyle[twocolumn,aps]{revtex}
\begin{document}
\draft
\title{ Two interacting particles in the Harper model
 }

\author { D.L.Shepelyansky$^{a,b}$}
\address {
Institute for Theoretical Physics,  University of California\\
Santa Barbara, CA 93106-4030
}
\date{ 31 May, 1996}
\maketitle

\begin{abstract}
Dynamics of two particles with short range repulsive or attractive
interaction is studied numerically in the Harper model.
It is  shown that interaction leads to appearance of 
localized states and pure-point spectrum component in the 
case when noninteractive system is quasi-diffusive
or ballistic. In the localized phase interaction gives only
stronger localization contrary to the case of two interacting
particles in a random potential.
\end{abstract}

The Harper model of electrons on a two-dimensional square lattice
in the presence of a perpendicular magnetic field was 
intensively studied during last decades (see e.g 
\cite{HAP,Aubry,HAP1,HAP2,HAP3,Last}). 
After fixing the quasimomentum in one of directions the
eigenvalue equation is reduced to a very simple form
of one-dimensional quasiperiodic discrete chain 
\begin{equation}
\begin{array}{c}
2\lambda \cos(\hbar n+\beta)\phi_{n} + \phi_{n+1}+\phi_{n-1}=E\phi_{n}
\end{array}
\end{equation}
where the effective Plank's constant $\hbar/2\pi$ gives 
the ratio of magnetic flux through the
lattice cell to one flux quantum, $\beta$ is a constant related to
quasimomentum. For the original problem 
of electrons in a magnetic field the parameter $\lambda$
should be fixed at $\lambda=1$ but generally one can consider
the model (1) at different values of $\lambda$.
Intensive analytical and numerical studies \cite{Aubry,Last,HAP1} 
for typical irrational values $\hbar/2\pi$ showed that
for $\lambda > 1$ the spectrum is
a pure-point like with gapes and all eigenstates are exponentially localized.
For $\lambda < 1$ the spectrum becomes continuous with extended eigenstates
corresponding to ballistic classical motion. For $\lambda=1$
the  situation is critical with singular-continuous multifractal
spectrum and power law localized eigenstates.

While the properties of one-particle Harper model are now well 
understood in many respects the question about
effects of particles interaction was not much investigated up to now. 
Indeed, the main physical problem of interacting particles at
finite particle density is very complicated both for analytical and
numerical investigations. One of the approaches to understanding of this 
problem is to analyze the effect of interaction of only two particles
in the Harper model. Recently such approach has given  a number of interesting
results for two interacting particles (TIP) in a random potential
showing that even repulsive particles can form a pair which
can propagate on a large distance \cite{TIP,Imry}. In this Letter I address
the problem of TIP in a quasiperiodic potential showing that 
here interaction effects can be quite different from the case of
random potential. The investigation of such type of model
will allow to analyze the stability of multifractal spectrum in respect to
interactions.

For two particles on a square lattice $(x,y)$ with magnetic flux and on site
interparticle  interaction the eigenvalue equation has the form:
\begin{equation}
\begin{array}{c}
e^{i\hbar y_1}\psi_{x_1 +1, y_1, x_2, y_2} + 
e^{-i\hbar y_1}\psi_{x_1 -1, y_1, x_2, y_2} +\\ 
\psi_{x_1, y_1 +1, x_2, y_2} + \psi_{x_1, y_1 -1, x_2, y_2} +\\ 
e^{i\hbar y_2}\psi_{x_1, y_1, x_2 +1, y_2} + 
e^{-i\hbar y_2}\psi_{x_1, y_1, x_2 -1, y_2} + \\
\psi_{x_1, y_1, x_2, y_2 +1} + \psi_{x_1, y_1, x_2, y_2 -1} + \\
\tilde{U} \delta_{x_1, x_2} \delta_{y_1, y_2} \psi_{x_1, y_1, x_2, y_2} 
= E \psi_{x_1, y_1, x_2, y_2} 
\end{array}
\end{equation}
Here $(x,y)$ are integers marking the sites of the square lattice,
the indices $1,2$ note two particles,  $\tilde{U}$ is on site interaction,
and $\hbar=2\pi\phi/\phi_0$ is determined by the ratio of 
magnetic flux $\phi$ through the unit cell to the quantum of flux $\phi_0$.
The direct investigation of equations (2) is a quite complicated problem.
Therefore, I reduce it to a simpler one with Bloch waves
propagating in $x-$direction 
$\psi_{x_1, y_1, x_2, y_2} = \varphi_{y_1, y_2}
\int dk_1dk_2 A_{k_1,k_2} \exp(i (k_1 x_1+k_2 x_2)) $
that leads to TIP in the Harper model 
with effectively renormalized interparticle interaction:
\begin{equation}
\begin{array}{c}
(2\lambda \cos(\hbar n_1+\beta_1)+
2\lambda \cos(\hbar n_2+\beta_2)+U\delta_{n_1, n_2})\varphi_{n_1, n_2} + \\
\varphi_{n_{1}+1, n_2}+\varphi_{n_{1}-1, n_2}+ 
\varphi_{n_{1}, n_{2}+1}+\varphi_{n_{1}, n_{2}-1}=E\varphi_{n_{1}, n_{2}}
\end{array}
\end{equation}
Here for generality there is introduced the parameter $\lambda$ which
should be taken equal to $1$ for the model (2) and 
the coordinates $y_{1,2}$ are replaced by $n_{1,2}$. 
The parameters $\beta_{1,2}$ are $\beta_{1,2}=k_{1,2}$ and the strength of
renormalized interaction is $U=\tilde{U} \int dk A_{k_1+k_2-k, k}/A_{k_1,k_2}$.
The investigation of the properties of the model (3) should
allow to understand better the properties of the original
TIP problem on $2d$-lattice (2). 

The time evolution of models (2), (3) is governed by eqs. (2), (3)
with $E$ in the right hand side replaced by $i \partial  / \partial t$. 
This evolution for the model (3) was studied numerically 
on effective 2-dimensional lattice with size $N \times N \leq 301 \times 301$.
The flux ratio was fixed at the golden mean value
$\hbar/2\pi = (\sqrt{5}-1)/2$.
For $\lambda =1$ the spectrum of noninteracting problem is 
singular-continuous with multifractal properties \cite{HAP1,HAP2,HAP3}.
Due to that the spreading over the lattice is similar to the diffusive
case with the second moments of probability distribution
$\sigma_{\pm} =<(n_1 \pm n_2)^2>$
growing approximately linearly with time (see Fig.1).
The switched on interaction leads to a significant
decrease in the rate of this growth, namely approximately
in 10 times for the case in Fig.1.
Here, initially at time $t=0$ two particles are 
located at the same site so that all probability
is concentrated at $n_1=n_2=0$ (the same initial
conditions were used in Figs.2-3). 
The analysis of the probability
distribution $P_{\pm}=\sum_{n_{\pm} = const} P(n_1,n_2)$
dependence on $n_{\pm}=|n_1 \pm n_2|/\sqrt{2}$ 
shows that its tail has Gaussian shape. However,
while in the noninteractive case all the probabilities spread
diffusively over the lattice in the case  with interaction a large
part  of probability (around $W_{loc} \approx 0.9$) remains 
localized in the vicinity of the initial position of particles 
within the interval $-5 \leq n_{1,2} \leq 5$ 
(Fig.2). 
The numerical simulations show
that in the interacting case the distribution mainly consists from two
parts. One of them represents localized states and is frozen near the
initial position of particles, another continues to diffuse as in
noninteractive case and corresponds to the Gaussian tail of 
distribution evolving in a diffusive way. Fig. 2 represents a typical
distribution shape at an instant moment of time. These numerical
data clearly demonstrate the qualitative change 
induced by interaction: appearence of localized component.

If initially the particles are located
on different sites then the value of $W_{loc}$ decreases with
growth of the initial distance $\Delta r$ between them but 
its value still remains quite large if initial distance
is about few sites.  For fixed $\Delta r$ the value of $W_{loc}$
is not sensitive to the initial choice of $n_{1,2}$ and $\beta_{1,2}$.
The dependence on time of probability to stay at the origin $P_0$
averaged over the time interval $[0,t]$ in shown in Fig.3.
Only in the noninteractive case $P_0$  goes to zero with 
time while for nonzero $U$ its value approaches to some constant.
It is interesting to note that asymptotically $P_0$ is larger than zero
not only for $\lambda=1$ but also for $\lambda < 1$ when
the noninteracting case has continuous spectrum 
with waves ballistically propagating along the lattice.
With the interaction decreasing the value of $P_0 $
decreases also but not very sharply (Fig.3),  also it is not sensitive to
the sign of $U$. In the localized phase $\lambda > 1$ the interaction
gives only a decrease of spreading over the lattice sites similarly 
to the case with $\lambda \geq 1$. For example, in the 
case of Fig.1 but with $\lambda = 1.05$ and $U=1$ the probability $P_0$
is approximately 9 times larger than for $\lambda=1.05$ and $U=0$.

The numerical results discussed above demonstrate that
the TIP behavior in a quasiperiodic potential is quite different
from the case of random potential \cite{TIP,Imry} where the
interaction produces mainly delocalizing effect.  The main reasons for this
difference are probably the following. 
The delocalization in the Harper model (1)
at $\lambda=1$ appears as the result of quantum tunneling between the sites
with closes energies $E_n =2 \lambda \cos(\hbar n +\beta)$
which are exponentially far from each other but are
close in energy. Apparently, the interaction destroys these tiny
resonance conditions that leads to appearance of localized states.
These states are localized in all directions on 2d lattice $(n_1,n_2)$
(see Fig.2). Therefore, they do not correspond to 
a situation in which interaction creats a coupled state which 
can propagate along the lattice. According to the numerical 
data these localized states are centered 
on the plane $(n_1,n_2)$ mainly along the
diagonal $n_1 = n_2$ (two particles are close to each other)
and their structure is approximately the same and independent of
the position along the diagonal. This means that these states
form a pure-point component in the energy spectrum.
The question how this component is placed in respect to 
the spectrum of noninteracting particles remains open.
One possibility is that this pure-point spectrum is located
completely outside of noninteractive band $[-8,8]$. 
Such a case in some sense would be similar to
an impurity state in a usual ballistic continuous band.
Another possibility is that the pure-point spectrum
is also partially located inside the band $[-8,8]$ in the
gaps which exist for noninteractive problem. The second
possibility looks to be more probable and more interesting. 
One of the indications
in this direction is that strong decrease of $U$ (Fig.3) does
not lead to disappearance of localized component.
It is naturally to assume that the structure of the pure-point 
spectrum remains approximately the same for $\lambda  < 1$
(al least for not very small values of $\lambda$).
According to numerical data the singular-continuous part
of the spectrum at $\lambda = 1$ is not 
completely destroyed by the interaction,
so that the quasi-diffusive spreading still takes place
(Fig. 1), but it would be quite desirable  to have rigorous
mathematical results about the structure of the spectrum
in the presence of interaction. 

The results discussed above were obtained for the TIP
in the one-dimensional Harper model (3).  They indicate
that the interaction induced localized states also
should exist in the original problem (2) of TIP on the
two-dimensional lattice with magnetic flux. Indeed,
here the interaction again should give a destruction
of tiny resonance conditions for tunneling. However, the
direct detailed investigations of the model (2) are
required to make definite conclusions, but numerical
studies of the model (2) are much more complicated than for the 
model (3). Finally, let us note that in the classical noninteractive
limit the dynamics of models (2), (3) is integrable.
Therefore, it would be interesting to study the effect of TIP
in the kicked Harper model \cite{KH} where the one-particle
classical dynamics is chaotic.

In conclusion, the numerical investigations of  TIP in the Harper
model (3) show that the attractive/repulsive interaction leads
to appearance of localized states and pure-point spectrum.
This happens in the case when noninteractive system $(U=0)$
has quasi-diffusive wave packet spreading with singular-continuous spectrum
$(\lambda = 1)$ or even 
for $\lambda < 1$ when it has the  ballistic wave packet propagation
and continuous spectrum. Such effect of interaction in quasiperiodic
systems is attributed to the interaction induced destruction 
of tiny resonance conditions which in noninteractive system allowed
to tunnel between quasiresonant states leading to infinite wave packet
spreading and decay of the probability to stay at the origin to zero.
In the localized phase interaction gives only a decrease of localization
length.  Therefore, the situation in the quasiperiodic
systems is quite different from the case of random potential
where interaction between two particles gives an increase of
localization length.

This research was supported in part by the National Science Foundation
under Grant No. PHY94-07194.

\begin{figure}
\caption{ Dependence of second moments 
$\sigma_{\pm}=<(n_1 \pm n_2)^2>$
on time $t$ for TIP in the Harper model (3) ($\sigma_+$ is full curve,
$\sigma_-$ is dashed). The parameters are  $\lambda =1$, $\hbar=\pi (5^{1/2}-1)$,
${\beta_{1,2}}=2^{1/2}$; $U=0$ for upper curves, $U=1$ for lower curves.
System size is $N \times N =301 \times 301$ sites, 
initially particles are at the same $\protect{n_{1,2}}=0$.}
\label{fig1}
\end{figure}
\begin{figure}
\caption{ Dependence of integrated
probability distributions $P_{+}$ 
(full curve) and $P_{-}$ (dashed) on $n^2_{\pm}=(n_1 \pm n_2)^2/2$
for the case of Fig.1 and $t=4000$. Left curves are for $U=1$,
right are for $U=0$ (shifted for clearity).}
\label{fig2}
\end{figure}
\begin{figure}
\caption{ Dependence of probability to stay at the initial state $P_0$
on time $t$ for: $\lambda = 1, U=1$; $\lambda=5/6, U=5/6$;
$\lambda=1, U=1/4$; $\lambda=1, U=0$ (curves from up to down)
and $\hbar, \beta_{1,2}$ are as in Fig.1.}
\label{fig3}
\end{figure}


\begin{references}
\bibitem[a]{byline1} On leave from  Laboratoire de Physique Quantique, 
UMR C5626 du CNRS, Universit\'{e} Paul Sabatier,
31062 Toulouse Cedex, France.
\bibitem[b]{byline2} Also Budker Institute of Nuclear Physics,
630090 Novosibirsk, Russia.
\bibitem{HAP} D.R.Hofstadter, Phys. Rev. B {\bf14}, 2239 (1976).
\bibitem{Aubry} S.Aubry and G.Andr\'e, Ann. Israel Phys. Soc.
{\bf 3}, 133 (1980).
\bibitem{HAP1} J.B.Sokoloff, Phys. Rep. {\bf 126}, 189 (1985).
\bibitem{HAP2} T.Geisel, R.Ketzmerick and G.Petschel,
Phys. Rev. Lett. {\bf 66}, 1651 (1991).
\bibitem{HAP3} M.Wilkinson and E.J.Austin, Phys. Rev. B {\bf 50}, 1420 (1994).
\bibitem{Last} Y.Last, Comm. Math. Phys. {\bf 164}, 421 (1994).
\bibitem{TIP} D.L.Shepelyansky, Phys. Rev. Lett. {\bf 73}, 2607 (1994).
\bibitem{Imry} Y.Imry, Europhys. Lett. {\bf 30}, 405 (1995).
\bibitem{KH} R.Artuso, F.Borgonovi, I.Guarneri, L.Rebuzzini and G.Casati,
Phys. Rev. Lett. {\bf 69}, 3302 (1992).
\end{references}
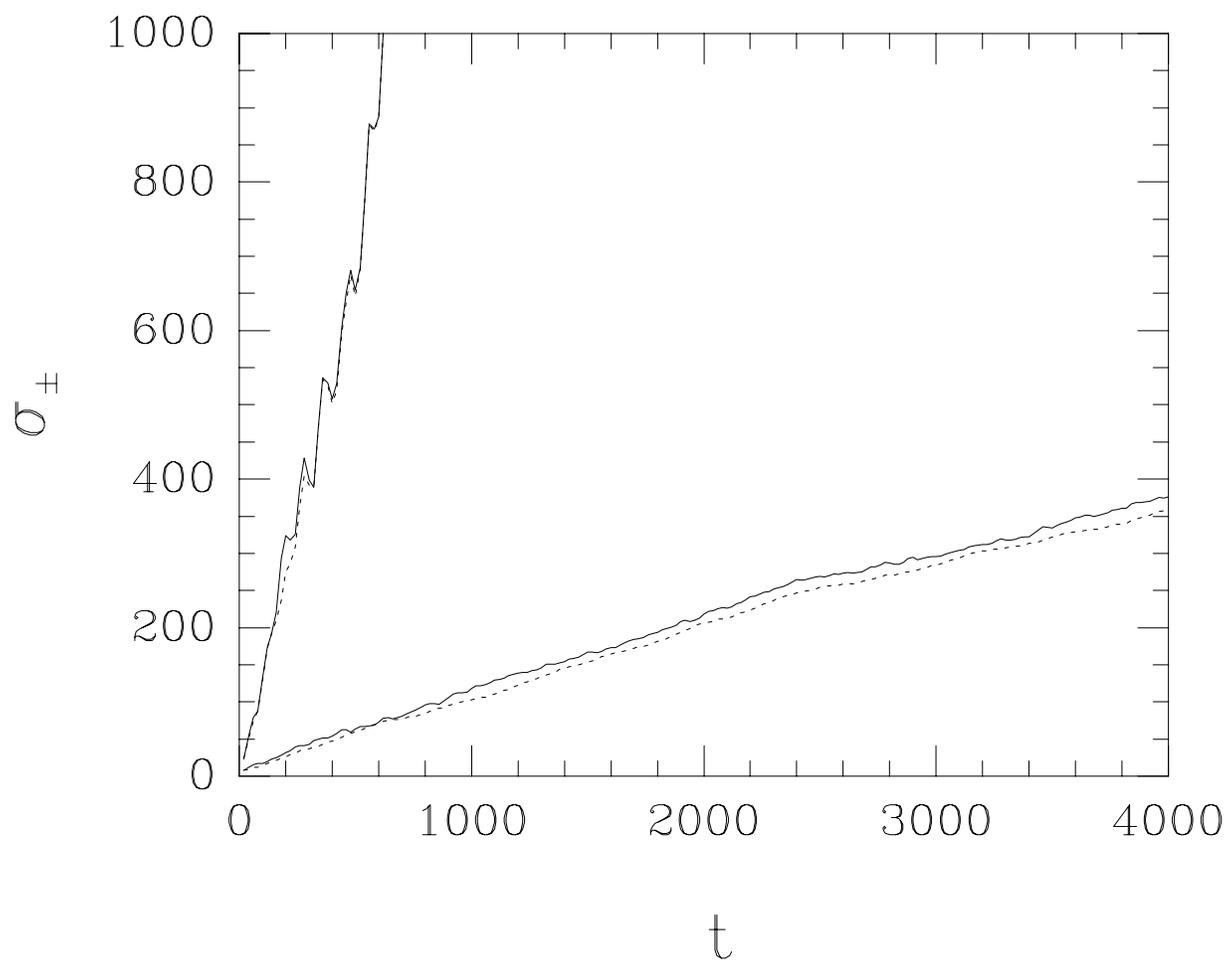
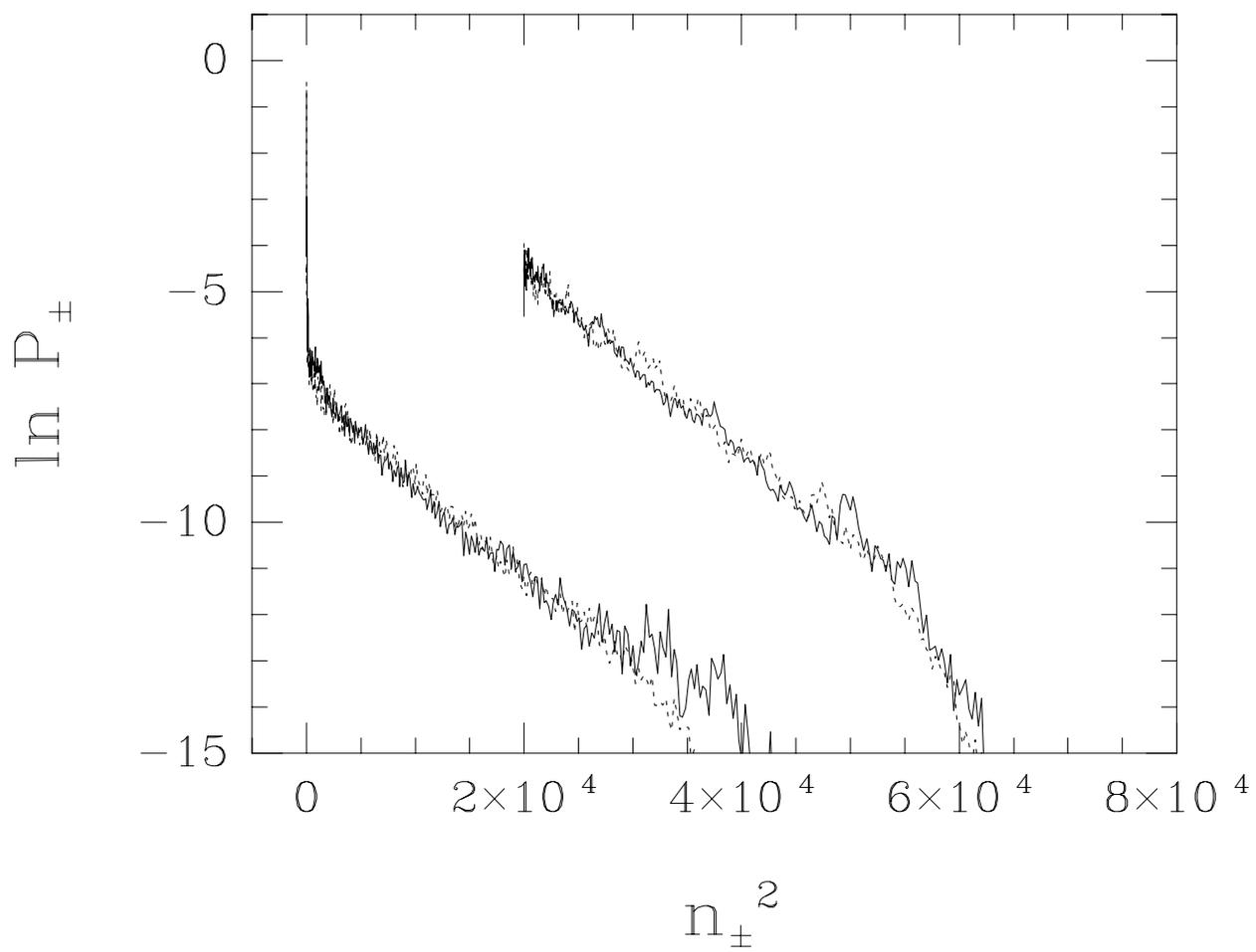
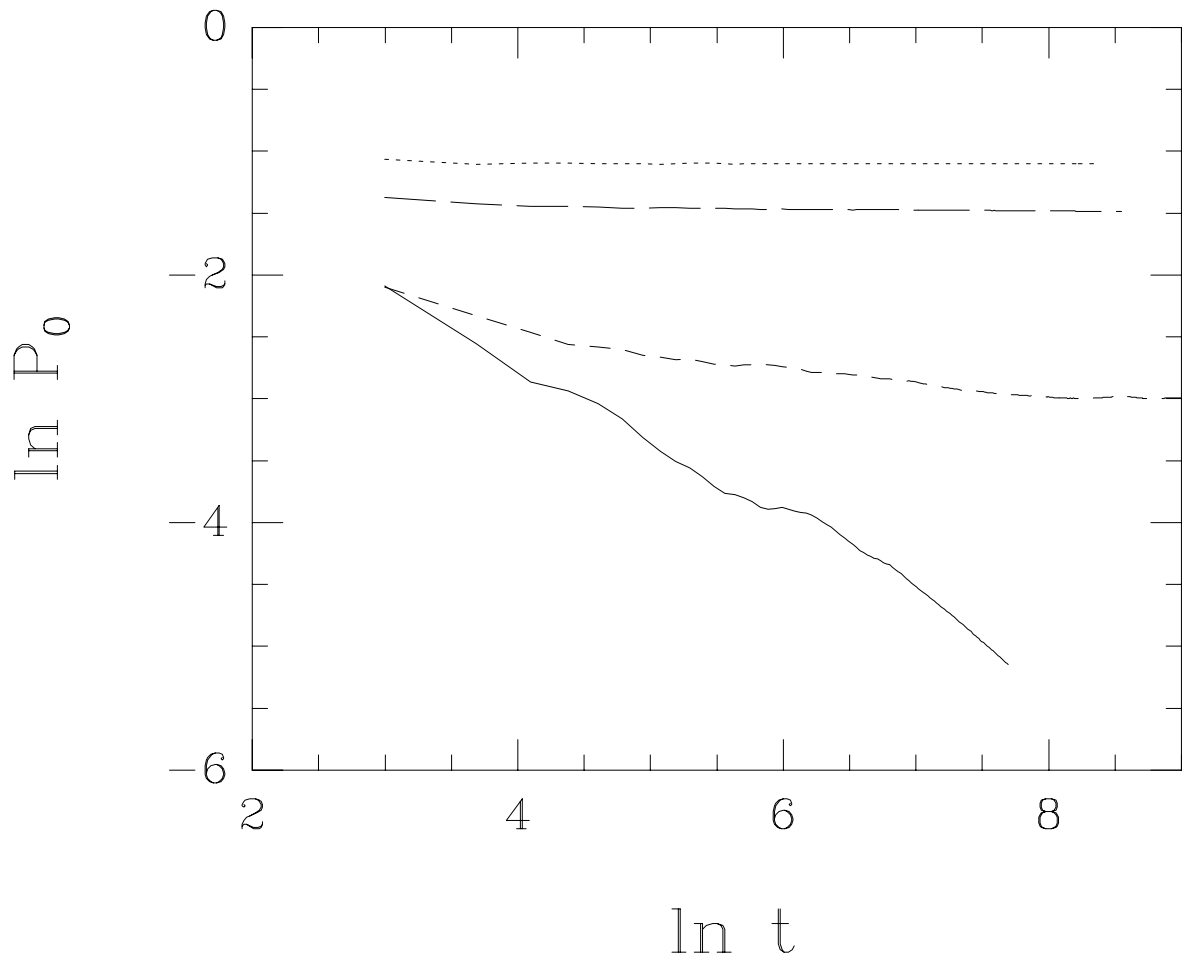
\end{document}